\documentclass[final,1p]{elsarticle}

\usepackage{graphicx}

\newcommand{\be}{\begin{equation}}
\newcommand{\ee}{  \end{equation}}
\newcommand{\ba}{\begin{eqnarray}}
\newcommand{\ea}{  \end{eqnarray}}
\newcommand{\ve}{\varepsilon}

\begin{document}

\begin{frontmatter}

\title{Spreading Widths of Doorway States}

\author[infn]{A. De Pace}
\author[dft,infn]{A. Molinari}
\author[mpi]{H.A. Weidenm\"uller}

\address[infn]{Istituto Nazionale di Fisica Nucleare, Sezione di Torino, 
  via P.Giuria 1, I-10125 Torino, Italy}
\address[dft]{Dipartimento di Fisica Teorica dell'Universit\`a di Torino,
  via P.Giuria 1, I-10125 Torino, Italy}
\address[mpi]{Max-Planck-Institut f\"ur Kernphysik, D-69029 Heidelberg, Germany}

\begin{abstract}
As a function of energy $E$, the average strength function
$\overline{S(E)}$ of a doorway state is commonly assumed to be
Lorentzian in shape and characterized by two parameters, the peak
energy $E_0$ and the spreading width $\Gamma^\downarrow$. The simple
picture is modified when the density of background states that couple
to the doorway state changes significantly in an energy interval of
size $\Gamma^\downarrow$. For that case we derive an approximate
analytical expression for $\overline{S(E)}$. We test our result
successfully against numerical simulations. Our result may have
important implications for shell--model calculations.
\end{abstract}


\end{frontmatter}

\section{Motivation}

Giant Resonances are an ubiquitous phenomenon in
nuclei~\cite{Bor98,Har01}. A specific nuclear mode with normalized
wave function $| 0 \rangle$ carrying definite quantum numbers (spin,
parity, isospin) is excited, for instance, by absorption of a gamma
quantum with specific multipolarity, by nucleon--nucleus scattering,
or by stripping of a nucleon from the projectile in the collision of
two nuclei. The mode has a typical mean excitation energy $E_0$ of
several or even 10 to 20 MeV, i.e., may occur above the first particle
threshold.  The Giant Dipole (GD) mode in nuclei is a paradigmatic
case. Aside from a normalization factor, the wave function $| 0
\rangle$ of the GD mode is the product of the dipole operator and the
eigenfunction of the nuclear ground state, and the dependence of $E_0$
on mass number $A$ is empirically given by $E_0 \approx 80 \ A^{-
1/3}$.  In general, the wave function $| 0 \rangle$ is not an
eigenstate of the nuclear Hamiltonian $H$ and the mode is, therefore,
not observed as a sharp and isolated resonance. Rather, the mode
spreads in a very short time $\tau^\downarrow$ (typically
$\tau^\downarrow \approx \hbar / 5$ MeV $\approx 2 \times 10^{-22}$
sec) over the eigenstates $| i \rangle$ of $H$ carrying the same
quantum numbers (each state $| i \rangle$ corresponding to an
eigenvalue $\ve_i$ of $H$). Thus, for the particular reaction under
consideration the mode $| 0 \rangle$ acts as a ``doorway'' to the
eigenstates of $H$ which manifests itself as a local enhancement of
the dependence on energy $E$ of the strength function \be S(E) =
\sum_i |\langle 0 | i \rangle|^2 \delta(E - \ve_i) \ .
\label{1}
\ee
The levels $\ve_i$ are actually particle--unstable and, thus,
resonances, and in most cases $S(E)$ is, therefore, a smooth function
of $E$. Often $S(E)$ displays a broad maximum. Pending the
modifications introduced below, the peak energy is then identified with
the mean excitation energy $E_0$ of the doorway state, and the width
$\Gamma^\downarrow$ is identified with $\hbar / \tau^\downarrow$ and
referred to as ``spreading width''. At excitation energies of $\approx
10$ MeV, the mean spacing $d$ of the nuclear levels $\ve_i$ is
typically of order $10$ eV, so that $\Gamma^\downarrow \gg d$. Hence
the name giant resonance. Similar phenomena also occur in
condensed--matter physics where the strength function $S(E)$ is
commonly referred to as the local density of states.

In the simplest theoretical model~\cite{BM68} for the giant--resonance
phenomenon, the doorway state $| 0 \rangle$ is coupled to a set of
background states $| \mu \rangle$ (where $\mu = 1, \ldots, N$ and $N
\to \infty$) via real coupling matrix elements $V_\mu$. The background
states have constant level spacing $d$. The matrix elements $V_\mu$
are Gaussian--distributed random variables with zero mean values and a
common variance $v^2$. The strength function is calculated in the
limit $N \to \infty$ as the average over the distribution of the
$V_\mu$ and given by~\cite{BM68}
\be
\overline{S(E)} = \frac{\Gamma^\downarrow / (2 \pi)}{(E - E_0)^2 +
(1/4) (\Gamma^\downarrow)^2} \ .
\label{2}
\ee
The bar denotes the ensemble average. The average strength function
has Lorentzian shape and is normalized to unity. The spreading width
is given by
\be
\Gamma^\downarrow = 2 \pi v^2 / d \ .
\label{3}
\ee
Although it looks like Fermi's golden rule, the result~(\ref{3}) is
correct beyond perturbation theory, i.e., for all values of $v^2 /
d^2$.

The level density $\rho(E)$ may be taken to be constant when the rate
of change with energy of $\rho(E)$ over an energy interval of length
$\Gamma^\downarrow$ is negligible, i.e., when $[{\rm d} \ln \rho(E) /
{\rm d} E]^{-1} \ll \Gamma^\downarrow$. In nuclei, that is not always
the case. By way of example we consider the GD mode in $^{16}$O. In
the shell model $| 0 \rangle$ is a superposition of one--particle
one--hole states. Through the residual interaction $| 0 \rangle$ is
coupled to two--particle two--hole states (two particles in the
$sd$--shell and two holes in the $p$--shell). The maximum spacing in
energy of the single--particle states (of the single--hole states) is
about $5$ MeV~\cite{Zel96} ($3$ MeV, respectively), giving the
spectrum of the two--particle two--hole states a spectral range of
about $15$ MeV. The residual interaction widens the range to $\approx
25$ MeV. The shape of the spectrum being Gaussian, the width $\sigma$
of the Gaussian is then around $15$ or $20$ MeV, and the ratio
$\Gamma^\downarrow \approx 5$ MeV to $\sigma$ is around $1/3$ or $1/4$
and, thus, not negligible. In the present paper we show how
Eq.~(\ref{2}) is modified under such circumstances.

Our investigation was triggered by a result for the strength function
of a doorway state obtained in Ref.~\cite{De07}. There we considered a
Hamiltonian matrix of the form
\be
H = \left( \matrix{ E_0 & V_\nu \cr V_\mu & {\cal H}_{\mu \nu} \cr}
\right) \ .
\label{4}
\ee
The doorway state $| 0 \rangle$ at energy $E_0$ is coupled to $N$
background states $\mu$ with $\mu = 1, \ldots, N$ and $N \to \infty$
via real matrix elements $V_\mu$. The background states are described
by a real--symmetric random Hamiltonian matrix ${\cal H}_{\mu \nu}$, a
member of the Gaussian Orthogonal Ensemble (GOE) of random matrices.
The average level density of ${\cal H}_{\mu \nu}$ has the shape of a
semicircle. Using the Pastur equation we calculated analytically the
average strength function (the ensemble average of $S(E)$ in
Eq.~(\ref{1})). Whenever the value of the spreading width
$\Gamma^\downarrow$ given by Eq.~(\ref{3}) was not negligible in
comparison to the radius $2 \lambda$ of the GOE semicircle, the
effective spreading width $\Gamma_{\rm eff}$ (defined as the full
width at half maximum of the average strength function) turned out to
be bigger than $\Gamma^\downarrow$, the increase being proportional to
$\Gamma^\downarrow / \lambda$. The method of derivation in
Ref.~\cite{De07} was confined to the GOE with its unrealistic
semicircular spectral shape. In the present paper we present an
approach that, although more approximate than that of
Ref.~\cite{De07}, applies for a coupling of the doorway state to
background states with a general dependence of the average level
density $\rho(E)$ on energy $E$. We determine how the effective
spreading width $\Gamma_{\rm eff}$ differs from $\Gamma^\downarrow$ as
given by Eq.~(\ref{3}) when $\rho(E)$ is not constant.

The model of Ref.~\cite{BM68} disregards all details of nuclear
structure. In a more realistic approach, one has to replace the
statistical assumptions on the matrix elements $V_\mu$ and the
assumption of a constant level spacing $d$ by a nuclear--structure
model like the shell model and/or one of the collective models. In
these approaches, the damping mechanism has received considerable
attention~\cite{Ber83,Bor98,Har01}, with special focus on the GD
resonance~\cite{Sar04}. Because of the large number of states that
couple to the doorway state, the effort is substantial, however, and
the simple statistical model of Ref.~\cite{BM68}, i.e., the use of
Eq.~(\ref{2}) together with a calculation of $\Gamma^\downarrow$ from
Eq.~(\ref{3}), continues to play an important role in the analysis of
giant--resonance phenomena in nuclei. For that reason we revisit and
extend the model in the present paper.

\section{Model}

Similarly to Eq.~(\ref{4}) we model the doorway state by the
Hamiltonian matrix
\be
H = \left( \matrix{ E_0 & V_\nu \cr V_\mu & E_\mu \delta_{\mu \nu} \cr}
\right)
\label{5}
\ee
where the index $\mu$ ranges from $1$ to $N$ with $N \gg 1$. The
matrix~(\ref{5}) differs formally from that of Eq.~(\ref{4}) in that
${\cal H}_{\mu \nu}$ has been diagonalized. Instead of the statistical
assumptions on the matrix ${\cal H}_{\mu \nu}$ made below
Eq.~(\ref{4}), we assume that the $V_\mu$ are Gaussian random
variables with zero mean value and a second moment $v^2$, and that
they are not correlated with the $E_\mu$. We do not need any
assumptions on the distribution of the latter. Thus, our model is more
general than the random--matrix model of Ref.~\cite{De07}.

To calculate the strength function $S(E)$, we rewrite Eq.~(\ref{1})
as
\be
S(E) = - \frac{1}{\pi} \ \Im \bigg( \langle 0 | \frac{1}{E^+ - H} | 0
\rangle \bigg)
\label{6}
\ee
where $E^+ = E + i \epsilon$ with $\epsilon$ positive infinitesimal.
Using Eq.~(\ref{5}) we obtain~\cite{Ber83,BM68}
\be
S(E) = - \frac{1}{\pi} \ \Im \bigg( \frac{1}{E^+ - E_0 - \sum_\mu
V_\mu (E^+ - E_\mu)^{-1} V_\mu} \bigg) \ .
\label{7}
\ee
Prior to calculating the ensemble average of $S(E)$ we calculate the
ensemble average of the sum over $\mu$ in the denominator of
Eq.~(\ref{7}). That sum is denoted by $\Sigma$. The average over the
distribution of the $V_\mu$ gives
\be
\overline{\Sigma}^V = v^2 \sum_\mu (E^+ - E_\mu)^{-1} \ .
\label{8}
\ee
For the remaining sum over $\mu$ we write
\be
F(E) = \sum_\mu \frac{1}{E^+ - E_\mu} = \int {\rm d} E' \ \frac{1}{E^+
- E'} \sum_\mu \delta(E' - E_\mu) \ .
\label{9}
\ee
Averaging over the distribution of the $E_\mu$, we replace $\sum_\mu
\delta(E' - E_\mu)$ by $\rho(E)$, the average level density of the
background states, and obtain
\ba
\overline{F(E)} &=& \int {\rm d} E' \ \frac{1}{E^+ - E'} \rho(E')
\nonumber \\
&=& - i \pi \rho(E) + \int {\rm d} E' \ \frac{\cal P}{E - E'}
\rho(E')
\label{10}
\ea
where ${\cal P}$ indicates the principal--value integral. Thus,
\be
\overline{\Sigma} = - i \pi v^2 \rho(E) + v^2 \int {\rm d} E' \
\frac{\cal P}{E - E'} \rho(E') \ .
\label{11}
\ee
We show presently that for $N \to \infty$ and $\Gamma^\downarrow \gg
d$ the average strength function $\overline{S(E)}$ is obtained by
replacing in Eq.~(\ref{7}) the function $\Sigma(E)$ by
$\overline{\Sigma}$. That yields
\be
\overline{S(E)} = \frac{1}{2 \pi} \ \frac{\Gamma^\downarrow}{(E - E_0
- \Delta)^2 + (1 / 4) (\Gamma^\downarrow)^2}
\label{12}
\ee
where
\ba
\Gamma^\downarrow &=& 2 \pi v^2 \rho(E) \ , \nonumber \\
\Delta &=& v^2 \int {\rm d} E' \ \frac{\cal P}{E - E'} \ \rho(E') \ .
\label{13}
\ea
Eqs.~(\ref{12}) and (\ref{13}) obviously generalize Eqs.~(\ref{2}) and
(\ref{3}) to the case where $\rho(E)$ is not constant and reduce to
the latter if it is.

To justify our averaging procedure (replacement of $\Sigma$ by
$\overline{\Sigma}$) we consider first the average of $S(E)$ over the
Gaussian--distributed matrix elements $V_\mu$. We use the property
that the average of the product $V_\mu V_\nu V_\rho V_\sigma$ has the
value $(v^2)^2 [\delta_{\mu \nu} \delta_{\rho \sigma} + \delta_{\mu
\rho} \delta_{\nu \sigma} + \delta_{\mu \sigma} \delta_{\nu \rho}]$,
and similarly for higher--order terms. In other words, averages over
products of Gaussian random variables are calculated by Wick
contraction of all pairs. For each pair the average is zero unless the
indices are equal. The average of $S(E)$ in Eq.~(\ref{7}) can be
calculated by expanding the denominator in powers of the $V$s and
using Wick contraction. After averaging, the leading contribution to
each term of the series is the one where $V$s appearing pairwise under
the same summation over $\mu$ are averaged. All other Wick
contractions restrict the independent summations over $\mu$ and lead
to terms that are small of order $1 / N$ and, thus, negligible for $N
\gg 1$. Hence to leading order in $1 / N$ averaging $S(E)$ in
Eq.~(\ref{7}) over the $V_\mu$ is equivalent to averaging $\Sigma$.

We turn to the average over the $E_\mu$ and use that $S(E)$ depends on
the $E_\mu$ only via the expression $\sum_\mu \delta(E - E_\mu)$.
Expanding $S(E)$ in powers of $\Sigma$ and averaging over the $E_\mu$
we see that our averaging procedure is justified if the $E_\mu$ are
uncorrelated. Then, in each term of the series $\sum_\mu \delta(E -
E_\mu)$ is replaced by $\rho(E)$ and the result is the same as
replacing $\Sigma$ in $S(E)$ by $\overline{\Sigma}$. The strongest
known correlations among eigenvalues are those of the GOE where the
$E_\mu$ follow Wigner--Dyson statistics. GOE level correlations extend
over an energy range measured in units of $d$ while $\overline{S(E)}$
varies with energy over an interval of length $\Gamma^\downarrow$.
Therefore, such correlations produce correction terms in the expansion
of $S(E)$ in powers of $\Sigma$ that are small of order $d /
\Gamma^\downarrow$ and are, thus, negligible for $\Gamma^\downarrow
\gg d$. The argument does not apply near the end points of the
spectrum where the level density tends to zero and $d$ becomes large.
This suggests that for our approximation to be valid the distance of
$E_0$ from the end points of the spectrum should be larger than
$\Gamma^\downarrow$. We observe, however, that equations~(\ref{12})
and (\ref{13}) provide reasonable approximations to our numerical
results even when that condition fails.

We conclude that Eqs.~(\ref{12}) and (\ref{13}) for the average
strength function $\overline{S(E)}$ of a doorway state are valid
except perhaps near the end points of the spectrum. These equations
generalize Eqs.~(\ref{2}) and (\ref{3}) by the appearance of a shift
function $\Delta(E)$. As shown by the second of Eqs.~(\ref{13}) that
function accounts for level repulsion between the doorway state and
the background states. The function $\Delta(E)$ receives negative
(positive) contributions from background states that lie above (below)
the energy $E$. If the spectrum is symmetric about $E = 0$ then
$\Delta(0) = 0$ and $\Delta(E) < 0$ ($\Delta(E) > 0$) if $E < 0$ ($E >
0$, respectively).  For a doorway state at $E_0 = 0$ this fact widens
the spectrum and causes $\Gamma_{\rm eff}$ to be larger than
$\Gamma^\downarrow$ as given by the first of Eqs.~(\ref{13}).
Obviously, our result agrees with Eqs.~(\ref{2}) and (\ref{3}) if the
average level density of the background states is constant so that
$\Delta = 0$. The shift function $\Delta(E)$ is very similar to the
shift function for a scattering resonance due to its interaction with
a continuum of scattering states~\cite{Ehr51,Tho52,Mit10}.

\begin{figure}[ht]
        \centering \includegraphics[clip,width=0.65\textwidth]{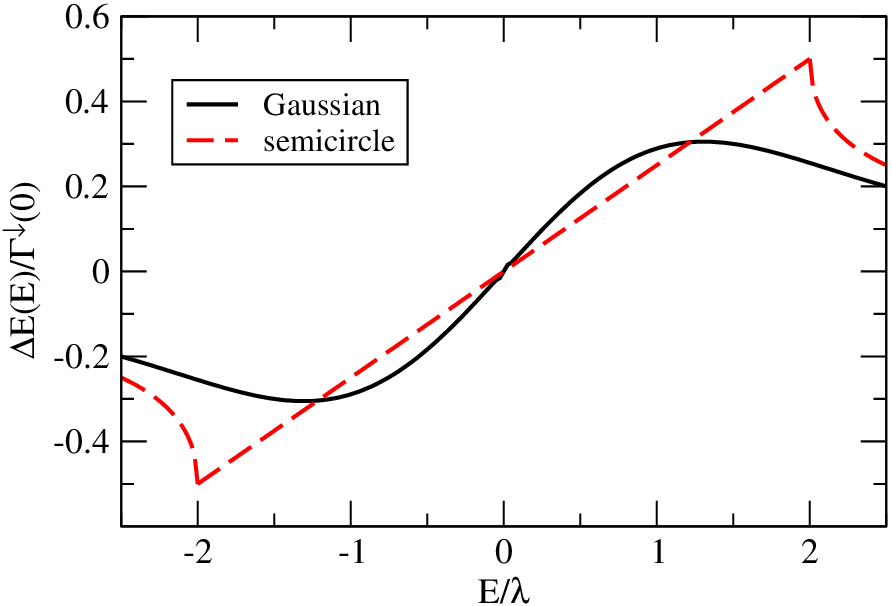}
        \caption{Plot of the functions $\Delta(E) /
        \Gamma^\downarrow(0)$ in Eqs.~(\ref{15}) versus $E/\lambda$
        for the semicircle (dashed, color online: red) and the Gaussian
        (solid, color online: black).}
\label{fig_Delta}
\end{figure}

We display the dimensionless ratio $\Delta(E) / \Gamma^\downarrow(0)$
for two important examples: The average level density has the shape of
a semicircle (the case of the GOE) or of a Gaussian (this is typical
of level densities in the shell--model~\cite{Zel96}). With the
normalization $\int {\rm d} E \ \rho(E) = N$ we have
\ba
\rho(E) &=& \frac{N}{\pi \lambda} \sqrt{1 - \bigg( \frac{E}{2 \lambda}
\bigg)^2} \ \ {\rm (semicircle)} \ , \nonumber \\
\rho(E) &=& \frac{N}{\sqrt{2 \pi} \lambda} \exp [ - E^2 / (2 \lambda^2)
] \ \ {\rm (Gaussian)} \ .
\label{14}
\ea
Here $\lambda$ denotes half the radius of the semicircle (the variance
of the Gaussian, respectively). With $x = E / \lambda$ the ratio
$\Delta(E) / \Gamma^\downarrow(0)$ is given by
\ba
\Delta(E) / \Gamma^\downarrow(0) &=& \frac{1}{2 \pi} \int_{-2}^{+2}
  {\rm d} x' \ \frac{\cal P}{x - x'} \sqrt{1 - x'^2/4} \nonumber \\
  &=& \frac{x}{4}-[\theta(2+x)-\theta(2-x)]\frac{1}{2}
    \sqrt{\frac{x^2}{4}-1}, \ \ {\rm (semicircle)}, \nonumber \\
\Delta(E) / \Gamma^\downarrow(0) &=& \frac{1}{2 \pi} \int_{- \infty}^{+
\infty} {\rm d} x' \ \frac{\cal P}{x - x'} \exp [ - x'^2 / 2 ] \nonumber \\ 
  &=& \frac{1}{2}{\rm e}^{-x^2/2}(-i){\rm erf}(i\frac{x}{\sqrt{2}})
  \ \ {\rm (Gaussian)}.
\label{15}
\ea
Fig.~\ref{fig_Delta} displays these two functions versus $x$. 
W conclude that $\Delta(E)$ is significant whenever $\Gamma^\downarrow
/\lambda$ is not negligibly small.

To estimate the effect of the values of $\Delta(E)$ displayed in
Fig.~\ref{fig_Delta} on the strength function, we note that for the
semicircle, the function $\Delta(E)$ is linear in $E$ over the entire
range of the spectrum while in the Gaussian case, $\Delta(E)$ is
approximately linear near the center of the spectrum. For both cases
we write 
\be
\Delta(E) \approx \alpha x \Gamma^\downarrow(0) = \alpha E \
(\Gamma^\downarrow(0) / \lambda) = \alpha E \gamma
\label{15a}
\ee
where $\gamma = \Gamma^\downarrow(0) / \lambda$ and where $\alpha > 0$
is dimensionless. Substituting that expression into Eq.~(\ref{12}) we
obtain
\be
\overline{S(E)} = \frac{1}{2 \pi} \ \frac{1}{1 - \alpha \gamma} \
\frac{\Gamma_{\rm eff}}{(E - \tilde{E}_0)^2 + \Gamma_{\rm eff}^2/4}
\label{16}
\ee
where 
\be
\tilde{E}_0 = E_0 / (1 - \alpha \gamma) \ {\rm and} \
\Gamma_{\rm eff} = \Gamma^\downarrow / (1 - \alpha
\gamma) \ .
\label{16a}
\ee
The factor $1 / (1 - \alpha \gamma) > 1$ shifts the mean energy $E_0$
of the doorway state towards smaller (larger) values when $E_0 < 0$
($E_0 > 0$, respectively) and increases the effective value of the
spreading width. The value of $\alpha$ is obtained by differentiating
$\Delta(E)$ at $E = 0$. For the semicircle we find $\alpha = 1/4$, in
agreement with the result of Ref.~\cite{De07}. For the Gaussian we
have $\alpha = 1 / \sqrt{2 \pi}$. In both cases and with $\gamma
\approx 1/3$ or $1/4$, that gives a correction of about $10$ to $20$
percent to both $E_0$ and $\Gamma^\downarrow$.

In summary, Eqs.~(\ref{12}) and (\ref{13}) are expected to provide a
better approximation to $\overline{S(E)}$ than Eqs.~(\ref{2}) and
(\ref{3}) if $\rho(E)$ changes significantly over an energy interval
of length $\Gamma^\downarrow$. Then we expect the full width
$\Gamma_{\rm eff}$ at half maximum of $\overline{S(E)}$ to be bigger
than $\Gamma^\downarrow$ as given by the first of Eqs.~(\ref{13}).
Eqs.~(\ref{12}) and (\ref{13}) may fail near the end points of the
spectrum of the background states. This is in accord with the exact
results of Ref.~\cite{De07}. There it was shown that the interaction
with the doorway state increases the range the GOE spectrum. Such an
effect is beyond the scope of the present approximate treatment.

\section{Numerical Simulation}

\begin{figure}[ht]
        \centering \includegraphics[clip,width=0.9\textwidth]{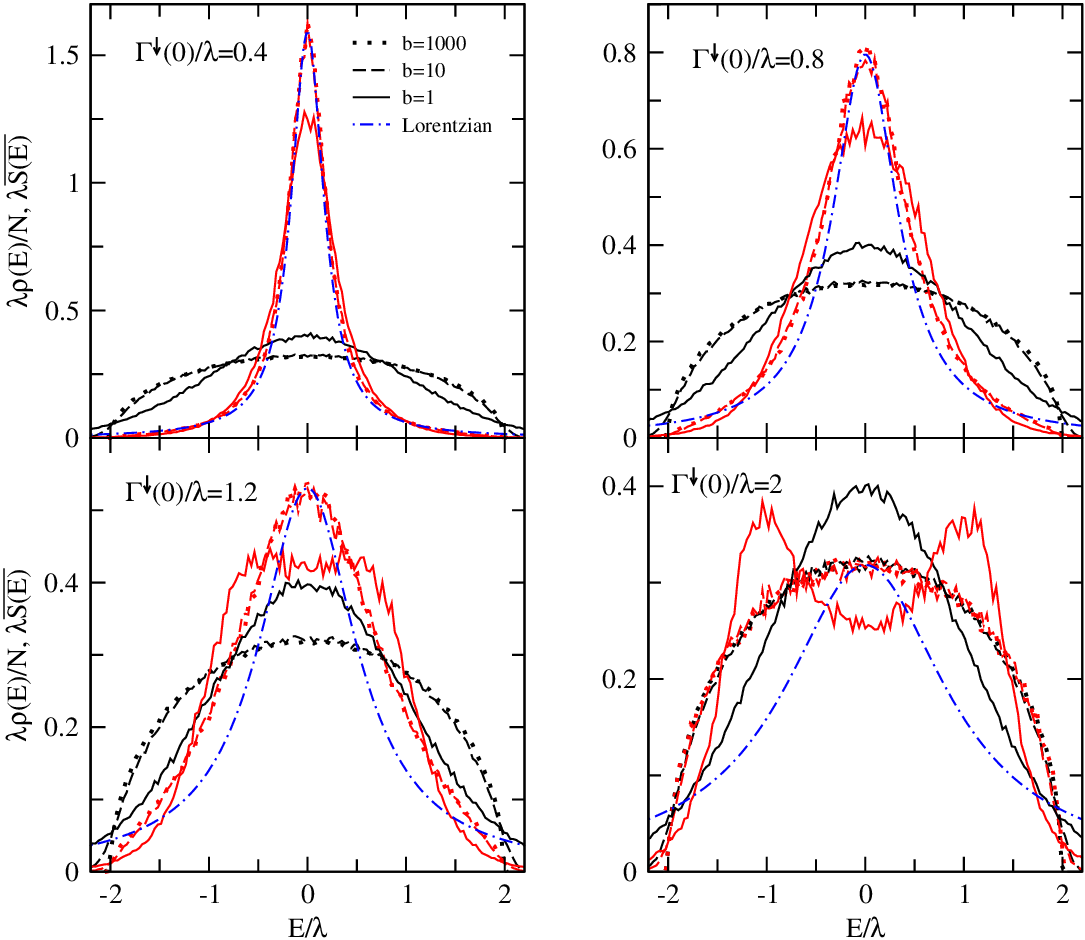} 
        \caption{Strength functions (color online: red) of a doorway
          state at $E_0 = 0$ coupled to background states described by
          a random band matrix of dimension $N = 1000$ and bandwidth
          $b$ as defined in the text; the coupling matrix elements are
          uncorrelated Gaussian-distributed random variables (case (i)
          of the text). For comparison, we also display the Lorentzian
          distribution of Eqs.~(\protect\ref{2}) and (\protect\ref{3})
          (color online: blue). The average level densities of the
          background states are also shown (color online: black). Each
          panel corresponds to a fixed value $\Gamma^\downarrow(0) /
          \lambda$ of the spreading width of Eq.~(\protect\ref{3}).
          Averages are performed over $m=500$ realizations.}
\label{fig_E0_0_GR}
\end{figure}

\begin{figure}[ht]
   \centering \includegraphics[clip,width=0.9\textwidth]{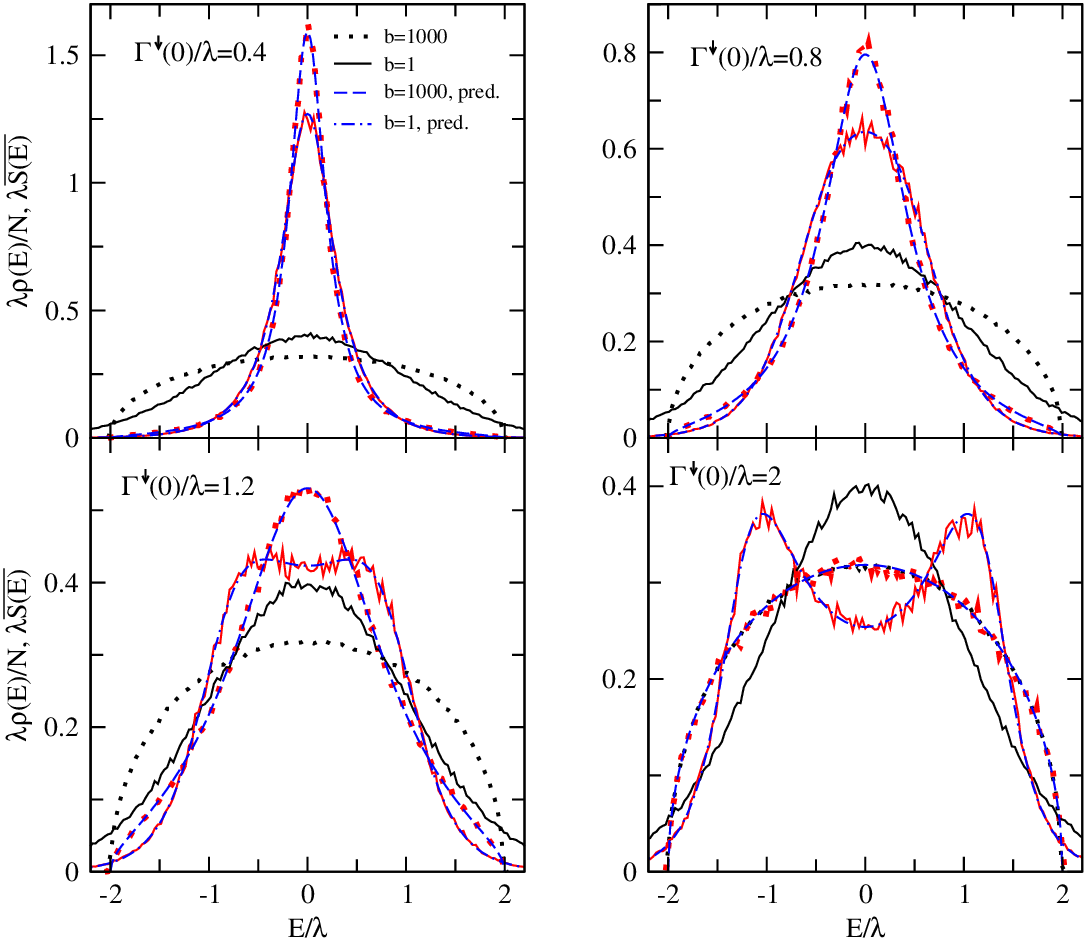} 
        \caption{Same as for Fig.~\ref{fig_E0_0_GR} but only for $b =
        1$ (diagonal random band matrix) and for $b = 1000$ (GOE
        matrix). In addition to the curves shown previously we now
        display also the predictions of Eqs.~(\protect\ref{12}) and
        (\protect\ref{13}) (color online: blue).}
\label{fig_E0_0_GR_ana}
\end{figure}

\begin{figure}[ht]
   \centering \includegraphics[clip,width=0.9\textwidth]{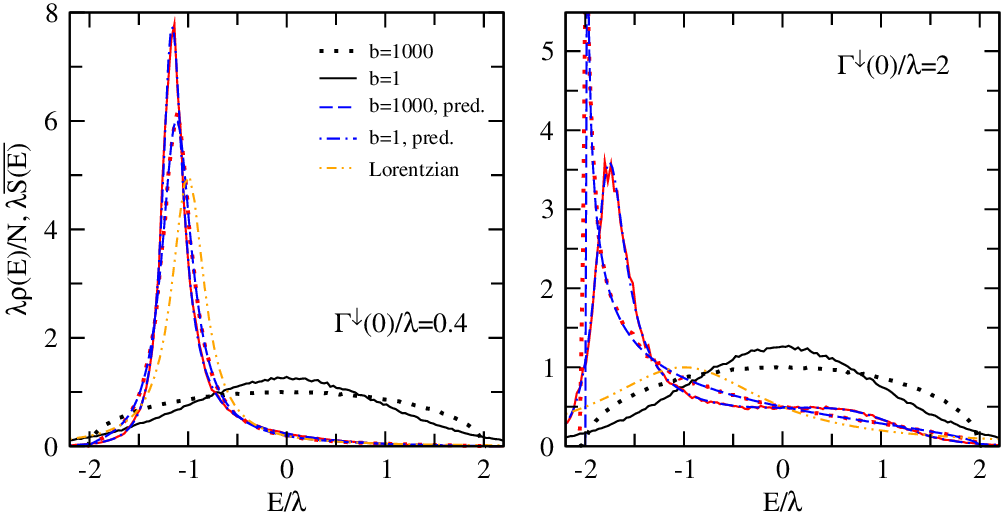} 
        \caption{Same as for Fig.~\ref{fig_E0_0_GR} at $E_0/\lambda=-1$, 
        but only for $b = 1$ (diagonal random band matrix) and for $b = 1000$
        (GOE matrix). We also display the predictions of
        Eqs.~(\protect\ref{12}) and (\protect\ref{13}) (color online: blue)
        and of Eqs.~(\protect\ref{2}) and (\protect\ref{3}) (color online:
        orange).} 
\label{fig_E0_1_GR}
\end{figure}

\begin{figure}[ht]
    \centering \includegraphics[clip,width=0.65\textwidth]{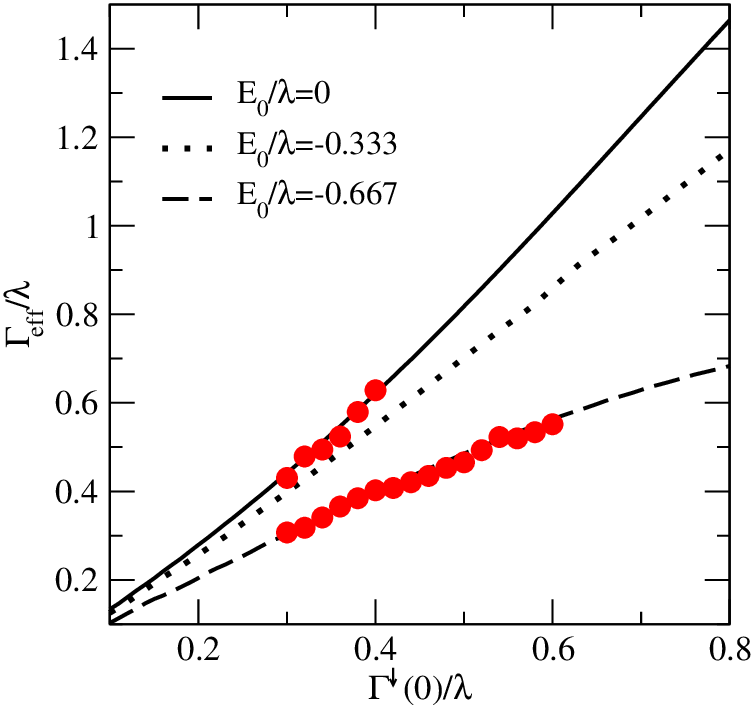} 
        \caption{Full width at half maximum $\Gamma_{\rm eff}$ of the
          average strength function of the doorway state, displayed as
          a function of the input width $\Gamma^\downarrow(0)$ of
          Eq.~(\ref{3}). The circles represent results of the
          numerical diagonalization, the lines correspond to
          Eqs.~(\protect\ref{12}) and (\protect\ref{13}).}
\label{fig_Geff}
\end{figure}

To test the approximations leading to Eqs.~(\ref{12}) and (\ref{13})
we consider a doorway state coupled to a random band matrix. Random
band matrices have been frequently used in different physical
contexts. The model is that of Eq.~(\ref{4}) except that ${\cal
H}_{\mu \nu}$ is a real symmetric random band matrix of dimension $N$:
All matrix elements with $|\mu - \nu| \ge b$ vanish. The upper bound on
the number of non--zero elements in every row and column is $(2 b -
1)$. The non--vanishing matrix elements are uncorrelated
Gaussian--distributed random variables with variances given by
$\overline{{\cal H}^2_{\mu \nu}} = (1 + \delta_{\mu \nu}) \beta^2$.
For $b = 1$ the matrix ${\cal H}_{\mu \nu}$ is diagonal while for $b =
N$ it is equal to the GOE. To make sure that all spectra have
approximately the same width we determine $\beta^2$ from the condition
$(1/N) {\rm Trace} \overline{{\cal H}^2} = \lambda^2$. That gives
\be
\beta^2 = \frac{\lambda^2}{2 b - \frac{\displaystyle b(b - 1)}
{\displaystyle N } }.
\label{17}
\ee
The average spectrum of the random band matrix ${\cal H}_{\mu \nu}$ is
Gaussian for $b = 1$ and changes quickly into an approximately
semicircular form as $b$ is increased~\cite{Cas90}. For $N = 1000$, we
found that the average spectrum is much more similar to a semicircle
than to a Gaussian already for $b = 5$; the transition to semicircular
shape was virtually complete at $b = 100$. For $b < \sqrt{N}$ and $N
\gg 1$ the eigenfunctions of a random band matrix are localized, and
the eigenvalues are uncorrelated, i.e., have Poissonian
statistics~\cite{Cas90}. Indeed, for $N = 1000$ the nearest--neighbor
spacing distribution changes from Poisson to Wigner form near $b =
30$. Similarly, the inverse participation ratio defined below
decreases strongly with increasing $b$. Some of these results are
displayed in the figures shown below. As a consequence, random band
matrices are useful for testing our approximations both for a Gaussian
spectrum ($b = 1$) and for a spectrum with Poisson statistics ($b <
\sqrt{N}$).

We have considered two ways of coupling the doorway state with the
random band matrix ${\cal H}_{\mu \nu}$: (i) The coupling matrix
elements $V_\mu$ in Eq.~(\ref{4}) are uncorrelated
Gaussian--distributed random variables with zero mean values and a
common second moment $v^2$. Then the doorway state is coupled to all
states in ${\cal H}_{\mu \nu}$ irrespective of the value of $b$, i.e.,
irrespective of the localization properties of the eigenvectors of
${\cal H}_{\mu \nu}$. (ii) We take $V_\mu = 0$ for all $\mu = 1,
\ldots, N$ except for $\mu$--values in a band of width $w$ centered in
the interval $[1, N]$. The doorway state is coupled only to select
states in ${\cal H}_{\mu \nu}$. Localization properties of the random
band matrix should influence the value of the average strength
function of the doorway state. In case (ii) the non--vanishing matrix
elements were taken to have all the same value $v \sqrt{N/w}$. Then
the total coupling strength $\sum_\mu V^2_\mu$ is on average the same
in cases (i) and (ii).

The input parameters of the model are $b, w, v^2, E_0$, and $N$ while
$\lambda$ defines the spectral width and, thus, the energy scale. A
further input parameter is $m$, the number of independent drawings of
the matrix elements ${\cal H}_{\mu \nu}$ from a random--number
generator. Each such drawing produces a realization of the random
matrix~(\ref{4}). Diagonalization of that matrix yields the
eigenvalues $\ve_i$ and eigenfunctions $| i \rangle$. These are used
to generate the strength function in Eq.~(\ref{1}). Combining $m$
realizations we obtain the average strength function
$\overline{S(E)}$.  That function is compared with Eqs.~(\ref{12}) and
(\ref{13}).

For case (ii) with parameter $w = 1$ the doorway state with energy
$E_0$ is coupled to only a single other state with
Gaussian--distributed energy $\ve$, and the average strength function
$\overline{S(E)}$ can be calculated analytically. We find
\be
\overline{S(E)} = \frac{1}{\sqrt{\pi} \lambda} \frac{v^2}{(E_0-E)^2}
\left.{\rm e}^{- \epsilon^2 / \lambda^2}\right|_{\epsilon = E
+ \frac{v^2} {(E_0-E)}} \ .
\label{18}
\ee
The function $\overline{S(E)}$ vanishes at $E = E_0$, extends over the
entire spectrum, and has two maxima on opposite sides of $E = E_0$.
That is a consequence of level repulsion and explains qualitatively
some of the features seen in the figures shown below.

In Figs.~\ref{fig_E0_0_GR} to \ref{fig_Geff} we present numerical
results for case (i). In Fig.~\ref{fig_E0_0_GR} we display strength
functions for parameter values indicated in the figure. The
discrepancy between the predictions of Eqs.~(\ref{2}) and (\ref{3})
and the actual values of the strength function are obvious and
increase with increasing values of the spreading width
$\Gamma^\downarrow$. We note the gradual development of a dip at $E_0
= 0$ and of a double--hump of the strength function. We believe that
these features correspond to properties of the simple model of
Eq.~(\ref{18}). In Fig.~\ref{fig_E0_0_GR_ana} we compare some of these
results with the predictions of Eqs.~(\ref{12}) and (\ref{13}) and
find very good agreement. We note that the dip is correctly
reproduced. We have performed similar calculations for non--zero
values of $E_0$ and found that the Lorentzian model shows even larger
discrepancies, since it is not able to reproduce not only the correct
width, but also the asymmetric shape that develops for $E_0\ne0$.  On
the other hand, Eqs.~(\ref{12}) and (\ref{13}) provide the same good
agreement for every value of $E_0$. An example is shown in
Fig.~\ref{fig_E0_1_GR} for the case of $E_0/\lambda=-1$, while our results are
summarized in Fig.~\ref{fig_Geff}. 
We believe that the agreement of the numerical
results with Eqs.~(\ref{12}) and (\ref{13}) is impressive. We also
note that $\Gamma_{\rm eff}$ and $\Gamma^\downarrow$ differ
significantly.

\begin{figure}[t]
     \centering \includegraphics[clip,width=0.9\textwidth]{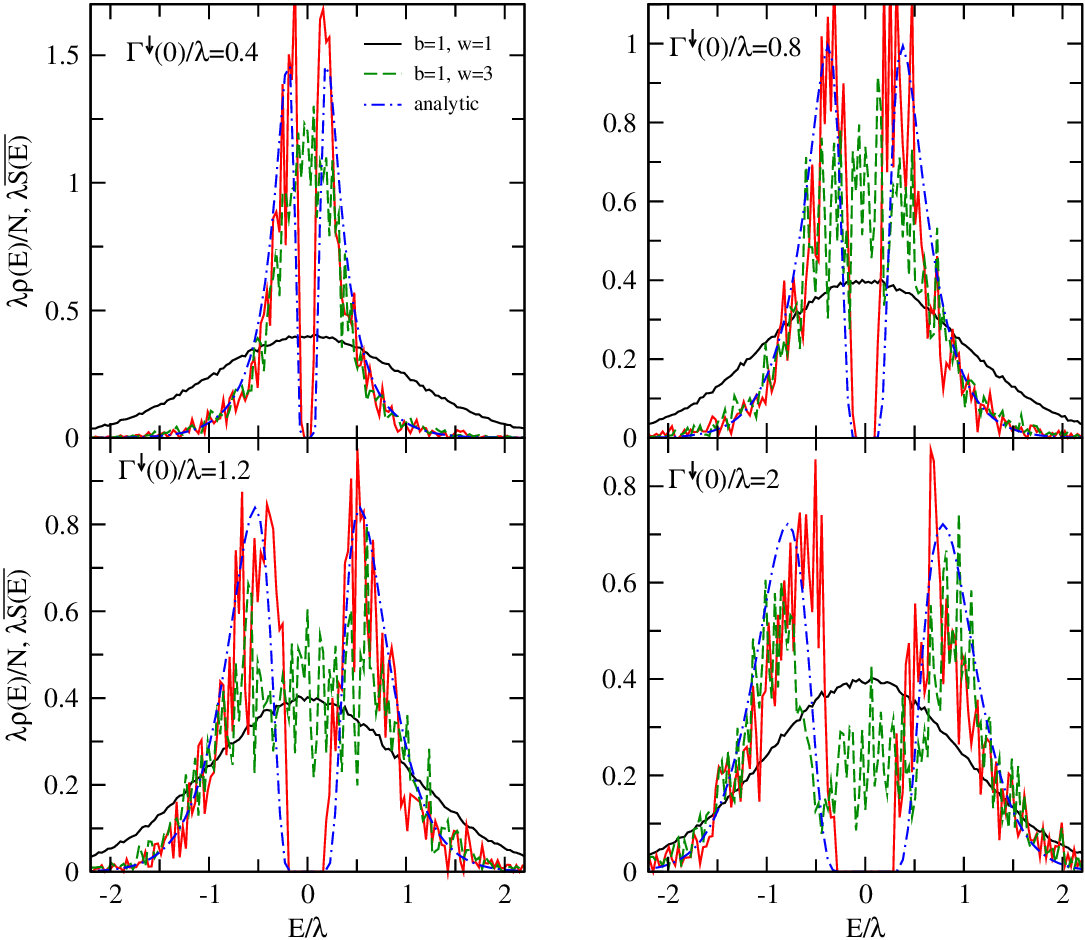} 
        \caption{The average strength function (color online: red) for
        a doorway state coupled to a random band matrix with $b = 1$
        of dimension $N = 1000$ (case (ii) of the text) is compared
        with the analytical result of Eq.~(\ref{18}) (color online:
        blue). The average level density of the backgound states is
        also shown (color online: black). The value of
        $\Gamma^\downarrow(0) / \lambda$ is given in each panel. The
        dashed lines (color online: green) show the strength function
        when the number of states directly coupled to the doorway
        state is increased from one to three.}
\label{fig_E0_0_BD_ana}
\end{figure}

\begin{figure}[ht]
    \centering \includegraphics[clip,width=0.9\textwidth]{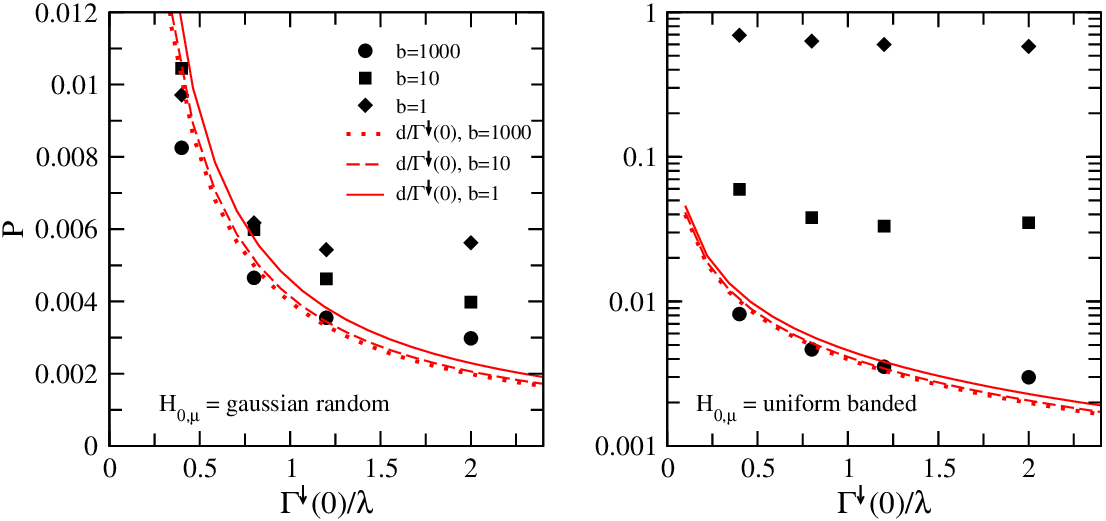} 
        \caption{Inverse participation ratio of the doorway state as a
          function of the input spreading width $\Gamma^\downarrow /
          \lambda$ for $N = 1000$ and several values of $b$ as shown
          in the panels. In the left panel the coupling matrix
          elements are uncorrelated Gaussian--distributed random
          variables (case (i) of the text), while in the right panel
          they are equal and restricted to a band of width $w = b$
          (case (ii) of the text). The average level spacing of each
          matrix ensemble is denoted by $d$.}
\label{fig_ipr_E0_0}
\end{figure}

We turn to case (ii). Again we have performed calculations for $N =
1000$. Case (ii) agrees with case (i) for $b = 1000$. The results for
$b = 10$ are qualitatively similar to those of case (i). Therefore, we
focus attention on the case $b = 1$ and consider $E_0 = 0$. Results
are shown and compared with the exact analytical result of
Eq.~(\ref{18}) in Fig.~\ref{fig_E0_0_BD_ana}. The double hump is
clearly displayed. The agreement is very good as expected. Similarly
good agreement was found when $E_0$ was chosen different from zero. On
the other hand, a comparison of Fig.~\ref{fig_E0_0_BD_ana} and the
results of Eqs.~(\ref{12}) and (\ref{13}) displayed in
Fig.~\ref{fig_E0_0_GR_ana} shows that the approximations leading to
Eqs.~(\ref{12}) and (\ref{13}) fail when the doorway state is coupled
to a single background state only (the case $w = 1$ in
Fig~\ref{fig_E0_0_BD_ana}). That is a very special situation and not
typical for doorway states. Increasing the number of background states
to which the doorway is coupled but keeping the bandwidth of ${\cal
H}_{\mu \nu}$ unchanged, very quickly changes the strength function so
that approximate agreement with Eqs.~(\ref{12}) and (\ref{13}) is
attained. For $w = 3$ that is shown by the dashed lines (color online:
green) in Fig.~\ref{fig_E0_0_BD_ana}.

To understand the role of localization in the mixing of the doorway
state with the background states we have calculated the average
inverse participation ratio (IPR) of the doorway state. The IPR is
defined in terms of the amplitudes $\langle 0 | i \rangle$ of the
expansion of the doorway state in the basis of eigenfunctions $| i
\rangle$, $i = 0, 1, \ldots, N$ of the matrix~(\ref{4}) as $\sum_i
\overline{|\langle 0 | i \rangle|^4}$. If the doorway state is spread
more or less uniformly over the eigenstates then the normalization
condition $\sum_i |\langle 0 | i \rangle|^2 = 1$ suggests that the IPR
has a value around $1 / N$. If, on the other hand, the doorway state
mixes with only a few of the eigenstates then the IPR should be much
larger than $1 / N$. Thus, for case (i) we expect values of the IPR
around $1 / N$ and for case (ii) much bigger values. The left panel of
Fig.~\ref{fig_ipr_E0_0} corresponds to case (i). The IPR (black dots)
decreases as $\Gamma^\downarrow$ increases. The values for $b = 1$ are
somewhat larger than those for $b = 1000$ but still close to $1 / N$.
The solid, dashed and dashed--dotted lines are obtained from the
simple estimate $d / \Gamma^\downarrow(0)$ for the IPR. For case (ii)
(right panel) and $b = 1$ and $b = 10$ the IPR is significantly larger
than $1 / N$ and roughly given by $1 / b$. That shows that the doorway
state mixes only with few ($\approx b$) states.

\section{Conclusions}

We have investigated the strength function of a doorway state coupled
to a number of background states in cases of strong coupling (the
spreading width $\Gamma^\downarrow$ is not small compared to the range
of the spectrum of background states). Our result in Eqs.~(\ref{12})
and (\ref{13}) generalizes the standard weak--coupling result and
agrees with it for weak coupling. We have tested our result by
numerical simulations. In most cases studied, we found perfect
agreement between the numerical results and Eqs.~(\ref{12}) and
(\ref{13}). Exceptions are found only when the doorway is coupled to a
single background state. That situation is atypical. Even when the
number of directly coupled states is increased from one to three,
approximate agreement with Eqs.~(\ref{12}) and (\ref{13}) is attained.

We have pointed out that the strong--coupling case is of practical
interest and may actually play a role in shell--model calculations.
That is the case whenever the spreading width $\Gamma^\downarrow$ is
not very small compared to the range in energy over which the average
level density of the background states changes significantly. Then our
Eqs.~(\ref{12}) and (\ref{13}) offer a more accurate description of
the strength function of the doorway state than do the standard
Eqs.~(\ref{2}) and (\ref{3}). Typically the full width at half maximum
$\Gamma_{\rm eff}$ of the average strength function is then larger
than the theoretical expression $\Gamma^\downarrow = 2 \pi v^2
\rho(E)$. The difference may be important for a comparison between
theory and data.

When the doorway state couples to a random band matrix with localized
eigenfunctions and when the coupling involves only a narrow band of
states (our case (ii)), we have found that the inverse participation
ratio of the doorway state is considerably larger than the inverse
matrix dimension. That shows that the doorway state mixes only with a
restricted number of localized states. The result is important for
practical applications. Indeed, in the nuclear shell model the doorway
state (a 1p 1h state) mixes with 2p 2h states which have a Gaussian
spectrum. But because of the presence of other modes of excitation,
the actual nuclear spectrum is not Gaussian in shape but increases
monotonically with energy, and one may ask what significance our
results have in view of this fact. However, the mixing of the 2p 2h
states with such other states is weak (otherwise shell structure would
not persist). Modeling such weak mixing in terms of a random band
matrix with localization, we have shown that our results remain valid
in the presence of other modes of excitation.

\end{document}